\documentstyle[12pt,epsf,epsfig,amsmath]{article}
\def\eqright #1\cr{\noalign{\hfill$\displaystyle{{}#1}$}}
\def\eqleft #1\cr{\noalign{\noindent$\displaystyle{{}#1}$\hfill}}

\def\oldreffmt#1{\rlap{[#1]} \hbox to 2\parindent{}}

\def\figfmt#1{\rlap{Figure {#1}} \hbox to 1in{}}

\def\sectioneq{\def\theequation{\thesection.\arabic{equation}}{\let
\holdsection=\section\def\section{\setcounter{equation}{0}\holdsection}}}%

\newcounter{holdequation}

\def\begineq #1\endeq{$$ \refstepcounter{equation}\eqalign{#1}\eqno
	(\theequation) $$}
\def\contlimit{\,{\hbox{$\longrightarrow$}\kern-1.8em\lower1ex
\hbox{${\scriptstyle (a\rightarrow0)}$}}\,}
\def\centeron#1#2{{\setbox0=\hbox{#1}\setbox1=\hbox{#2}\ifdim
\wd1>\wd0\kern.5\wd1\kern-.5\wd0\fi
\copy0\kern-.5\wd0\kern-.5\wd1\copy1\ifdim\wd0>\wd1
\kern.5\wd0\kern-.5\wd1\fi}}
\def\centerover#1#2{\centeron{#1}{\setbox0=\hbox{#1}\setbox
1=\hbox{#2}\raise\ht0\hbox{\raise\dp1\hbox{\copy1}}}}
\def\centerunder#1#2{\centeron{#1}{\setbox0=\hbox{#1}\setbox
1=\hbox{#2}\lower\dp0\hbox{\lower\ht1\hbox{\copy1}}}}
\def\lsim{\;\centeron{\raise.35ex\hbox{$<$}}{\lower.65ex\hbox
{$\sim$}}\;}
\def\gsim{\;\centeron{\raise.35ex\hbox{$>$}}{\lower.65ex\hbox
{$\sim$}}\;}

\def\super#1{\ifmmode \hbox{\textsuper{#1}}\else\textsuper{#1}\fi}
\def\textsuper#1{\newcount\holdspacefactor\holdspacefactor=\spacefactor
$^{#1}$\spacefactor=\holdspacefactor}

\def\getcite#1,{\advance\citenumber by1
\def\getcitearg{#1}\def\lastarg{@}
\ifnum\citenumber=1
\ref{#1}\let\next=\getcite\else\ifx\getcitearg\lastarg\let\next=\relax
\else ,\ref{#1}\let\next=\getcite\fi\fi\next}

\def\pom{{\rm P\kern -0.53em\llap I\,}}
\def\spom{{\rm P\kern -0.36em\llap \small I\,}}
\def\sspom{{\rm P\kern -0.33em\llap \footnotesize I\,}}

\relax
\def\contlimit{\,{\hbox{$\longrightarrow$}\kern-1.8em\lower1ex
\hbox{${\scriptstyle (a\rightarrow0)}$}}\,}
\def\upon #1/#2 {{\textstyle{#1\over #2}}}
\relax
\renewcommand{\thefootnote}{\fnsymbol{footnote}}

\sectioneq

\def\til#1{\centeron{\hbox{$#1$}}{\lower 2ex\hbox{$\char'176$}}}
\def\tild#1{\centeron{\hbox{$\,#1$}}{\lower 2.5ex\hbox{$\char'176$}}}
\def\sumtil{\centeron{\hbox{$\displaystyle\sum$}}{\lower
-1.5ex\hbox{$\widetilde{\phantom{xx}}$}}}

\begin{document} 

\begin{titlepage} 

\rightline{\vbox{\halign{&#\hfil\cr
&\today\cr}}} 
\vspace{0.25in} 

\begin{center} 
  
{\large\bf Sextet Quarks and the Pomeron at the LHC}\footnote{Work 
supported by the U.S.
Department of Energy under Contract
W-31-109-ENG-38} 

\medskip

Alan. R. White\footnote{arw@hep.anl.gov }

\vskip 0.6cm

\centerline{Argonne National Laboratory}
\centerline{9700 South Cass, Il 60439, USA.}
\vspace{0.5cm}

\end{center}

\begin{abstract}
 
Adding two color sextet quarks to QCD gives many special features.
The high-energy S-Matrix, constructed via reggeon 
diagrams and chiral anomalies, contains the Critical Pomeron and
electroweak symmetry breaking is produced, by sextet pions.
Cosmic ray phenomena suggest large cross-section effects will be seen
at the LHC, in particular, involving the pomeron.
The sextet sector  embeds, uniquely, in a massless,
confining, left-handed SU(5) theory. The 
anomaly based high-energy S-Matrix could be that of the  
full Standard Model.

\end{abstract} 

\vspace{2in}

\centerline{Presented at the XIth Blois Conference on 
Elastic and Diffractive Scattering, Blois, France, May, 2005}

\renewcommand{\thefootnote}{\arabic{footnote}} \end{titlepage}

\section{High Energy $QCD_S$ } 

When two, massless, color sextet quarks are 
added to QCD (with six, massless, triplet quarks already present)
we obtain~\cite{arw04}
a special version of QCD that we call $QCD_S$. 
Particular infra-red and ultra-violet properties of QCD$_S$
allow the high-energy behavior to be constructed via the  
reggeon diagrams of ``$~CSQCD_S$'', in which SU(3) color is broken to SU(2).
``Non-perturbative'' amplitudes appear via  
chiral anomalies occuring in reggeon effective vertices
that contain triangle 
\begin{figure}[ht]
\centerline{\epsfxsize=5.8in\epsfbox{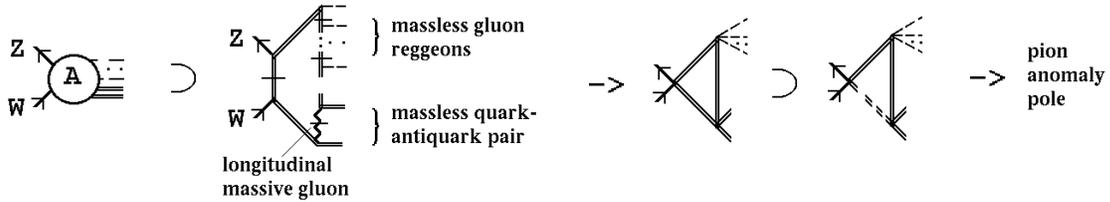}}
\caption{A reggeon vertex triangle diagram.}
\end{figure}
diagrams, generated as illustrated in Figure 1.
With a $k_{\perp}$ cut-off, vector Ward identities 
are violated for the anomaly vertices and ``wee gluon''
infra-red divergences occur coupled to anomaly poles.
The divergence of  
color zero, anomalous color parity ($C\neq \tau$) gluons is
preserved to all orders. Factorizing off the divergence as a ``wee
gluon condensate'' Goldstone 
boson anomaly pole ``pions'' are selected as physical states and 
the simplest $~\pi - \pi~$ scattering diagrams have the form shown in Figure 2.
Within each anomaly vertex there is a zero momentum chirality
transition or,
equivalently, a Dirac sea shift.  
The exchanged pomeron is a massive (reggeized) gluon in a wee gluon condensate
(with, in higher-orders, supercritical RFT interactions.)

SU(3) color is restored via the Critical Pomeron  
phase transition (asymptotic freedom properties of $CSQCD_S$ are crucial).
The wee gluon condensate disappears
and the Dirac sea shifting becomes dynamical. 
The physical states all originate as Goldstone bosons in $~CSQCD_S$.   
There are triplet mesons and nucleons, 
sextet ``pions''  ($\Pi$'s) and ``nucleons'' ~($P_6$ and $N_6$). There are no
hybrid sextet/triplet states and no glueballs. There is also no BFKL pomeron
and no  odderon.
\begin{figure}[ht]
\centerline{$~~~~~~$ \epsfxsize=3in
\epsffile{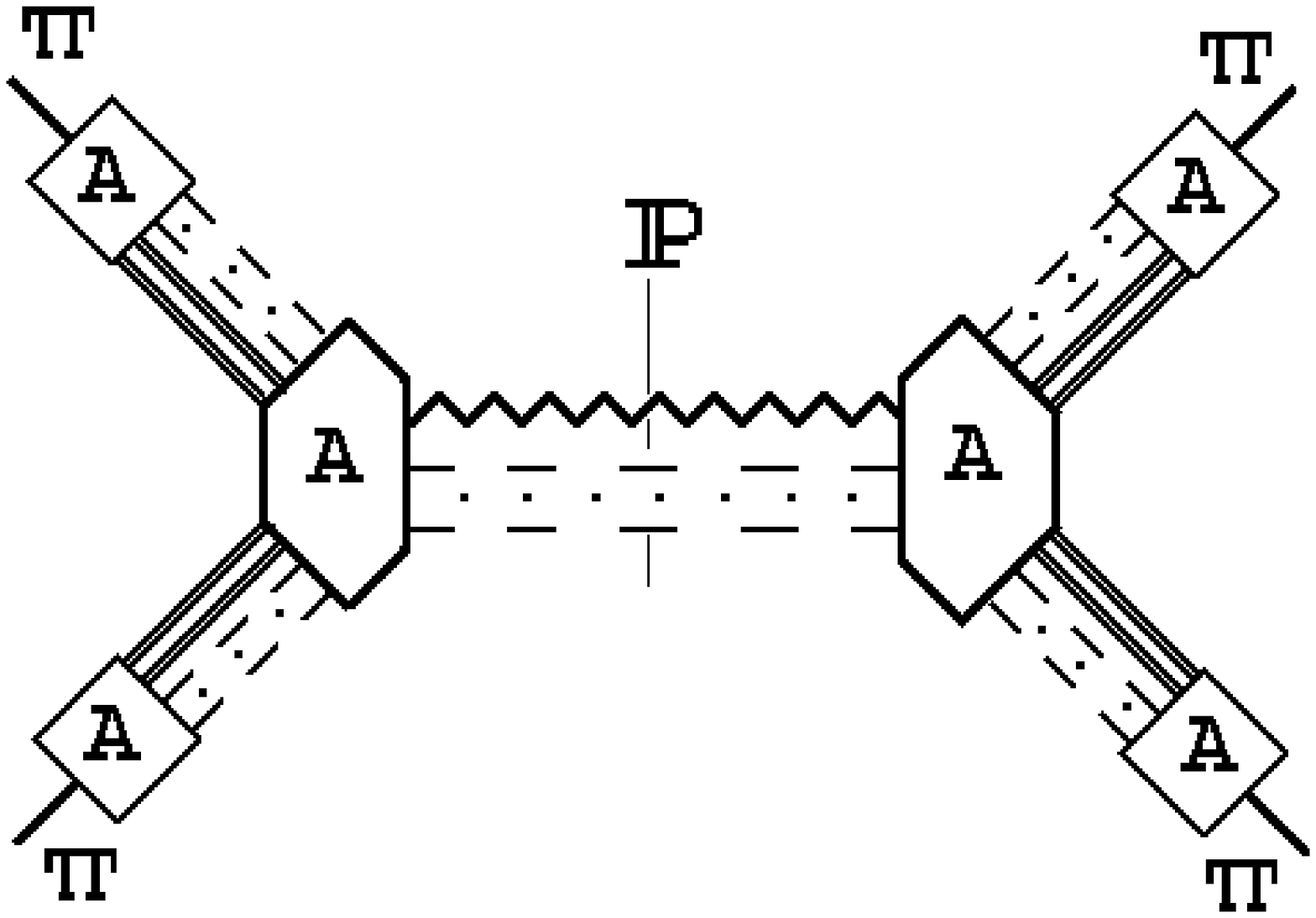}
\hspace{0.4in}
\epsfxsize=2in
\epsffile{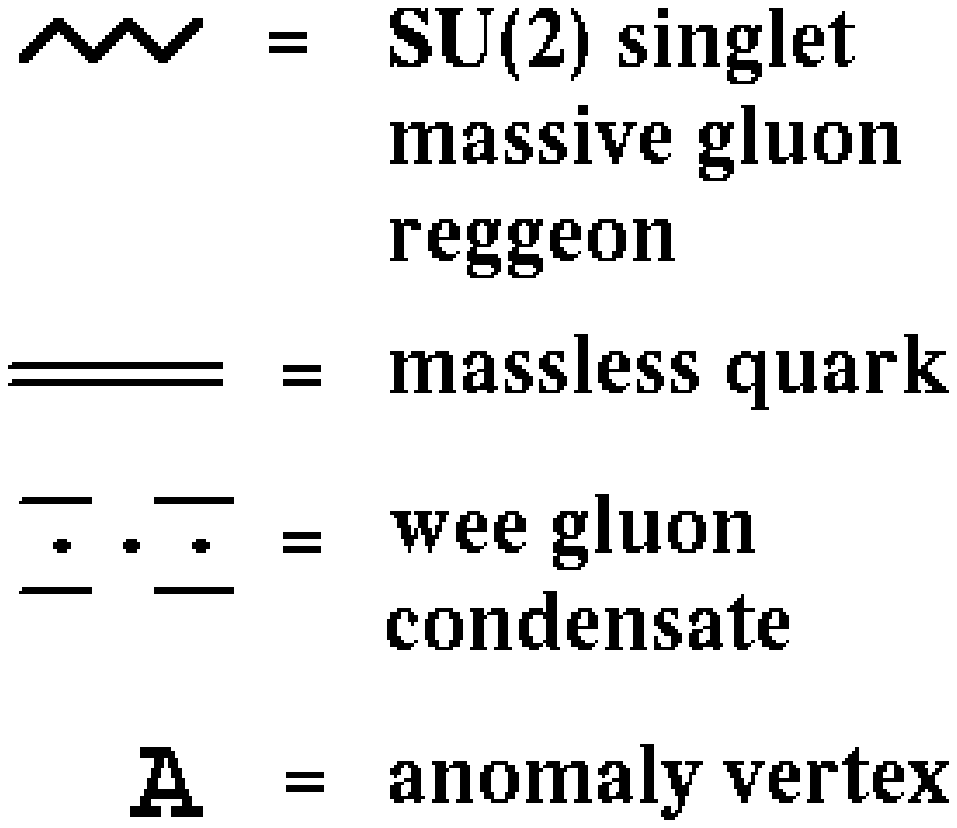}}
\caption{Pion Scattering}
\end{figure}

\section{The Sextet QCD Scale, Electroweak Masses, and Existing Evidence}

Wee gluons in the scattering states produce 
interactions that mix anomaly poles with exchanged electroweak vector bosons
and generate a mass.
This mass appears only in the S-Matrix and only 
for vectors with a left-handed coupling. No photon (or gluon) mass 
is generated. 
For a sextet quark loop $M_W^2 ~\sim~ g_W^2 ~\int dk ~k ~~\equiv ~ 
g_W^2 ~F_{\Pi}^2~$ where $k$ is a wee gluon momentum. 
Sextet quarks dominate because of larger color factors and 
the Casimir Scaling rule ($
C_6~\alpha_s (F_{\Pi}^2)~\sim ~C_3 ~\alpha_s(F_{\pi}^2)$ with $C_6/C_3 ~\approx~ 3$)
implies that $F_{\Pi}$ can consistently be the electroweak scale.
The large wee gluon coupling implies the pomeron
couples very strongly ($~\sim F_{\Pi}~$) to sextet quarks. 

Evidence for the sextet sector may, perhaps, 
have already been seen at HERA
and Fermilab. An anomaly pole  $\Pi$ can be produced via a  
large $k_{\perp}$ ``hard interaction'' 
of the pomeron with a color neutral $ \gamma$, $Z^0$, or $W^{\pm}$.
The diffractive DIS amplitude
is strongly enhanced by the anomaly
when $k_{\perp}$ is electroweak scale, and so,
the largest $~Q^2~$ ($> 40,000 ~GeV^2$)~
event seen by ZEUS  (before 97) could be diffractive production of
a $Z^0$. Different methods for 
reconstruction of the event give significantly different results 
(outside the errors) and the disagreement could result from a large jet mass. 
Consistency is achieved if the jet has a mass squared of
$8,077~GeV^2~=~ (89.9~GeV)^2 $
suggesting that a massive $Z^0$ jet was indeed produced.

Diffractive (sextet pion) hard vertices will also appear  
when a $Z^0$, $W^{\pm}$, or $\gamma$, is emitted from  
a quark in a hadron, but cross-sections will be relatively 
small and not easy to detect. Diffractive production
of a $W$ or $Z$ pair via a double anomaly pole vertex may 
give a detectable cross-section. An 
excess $W^+W^-$ cross-section (2 events) was apparently 
observed at the $S\bar{p}pS$ by UA1 and should, presumably, be also seen at the Tevatron.
Double pomeron production of $W^+W^-$ is probably 
not observable below LHC energies, but $\gamma Z^0$ might be seen.

The $\eta_6~$ mixes with a pure glue state and could be responsible for
$t\bar{t}$ production. If so, $m_t$ is a sextet scale and  
$\alpha_s$ evolution should stop at $E_T \sim m_t ~$,
giving a jet excess just as in Figure 3. 
\begin{figure}[ht]
\centerline{ a) 
\epsfxsize=2.5in
\epsfysize=1.5in
\epsfbox{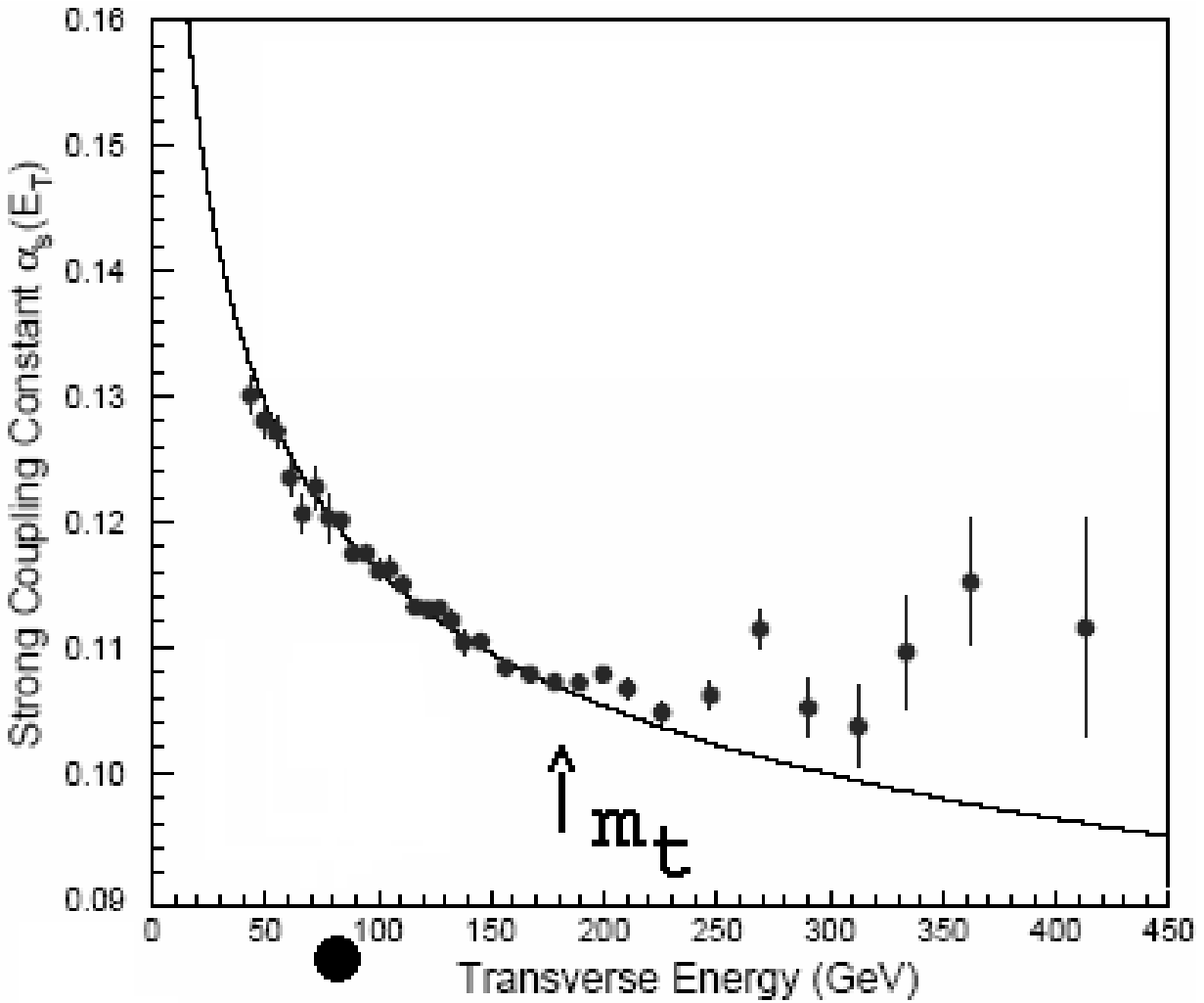}
\hspace{0.4in} b) 
\epsfysize=1.5in
\epsfbox{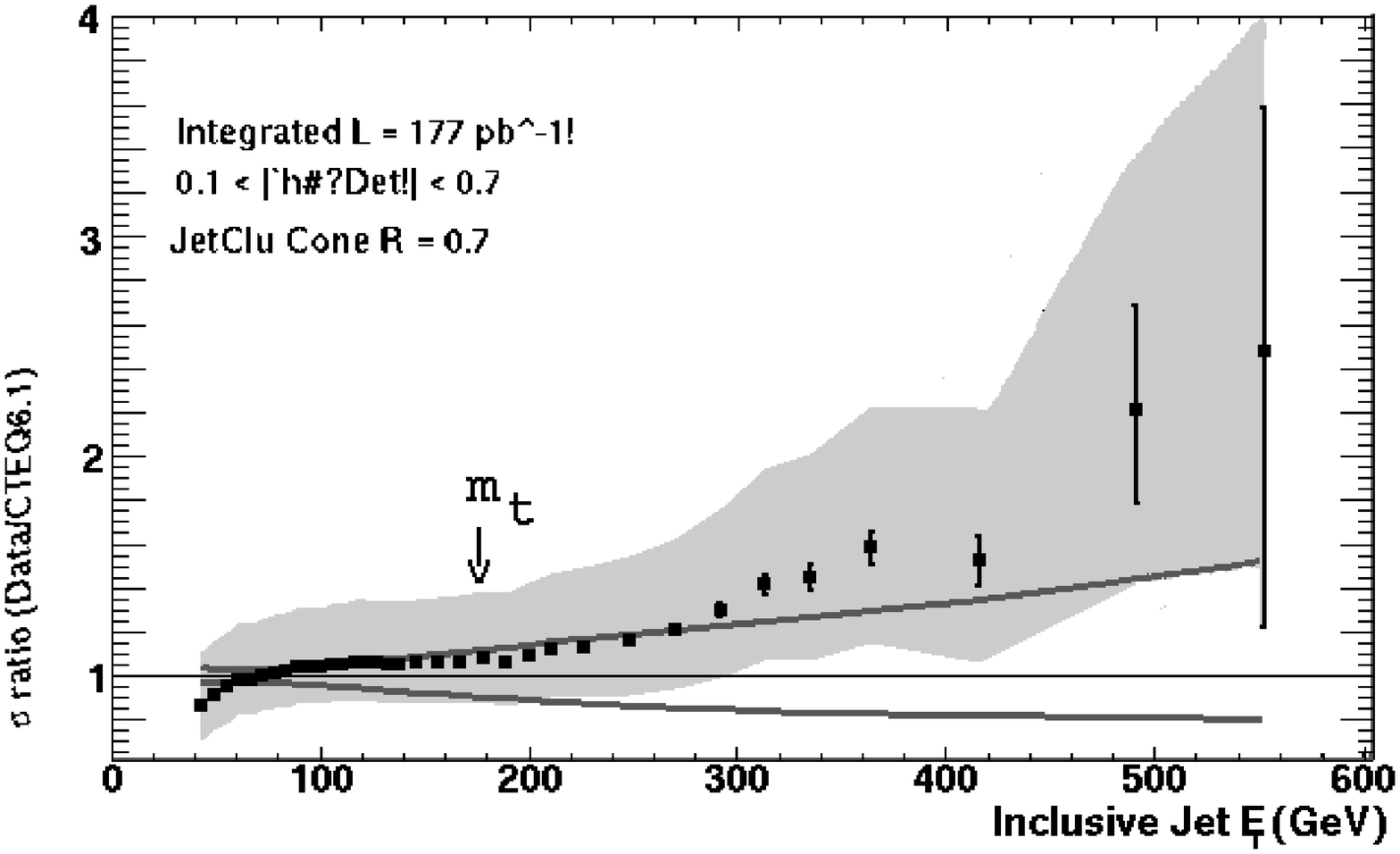}}
\caption{ CDF measurements of a) $\alpha_s$ in Run I and b) the 
jet cross-section in Run II}
\end{figure}

\section{Dark Matter and Cosmic Ray Physics }

Because of the Casimir effect, the sextet sector will constitute
a stronger coupling sector of $QCD_S$. Inclusive pomeron amplitudes will give the 
largest effect and, above an effective threshold, sextet states will start to 
dominate the inelastic x-section. In particular multiple $W^{\pm}$ and $Z^0$
production will cover most of the rapidity axis - in analogy
with low-energy pion production.
Because of the absence of hybrid triplet/sextet states, 
the sextet neutron (the $N_6$) will be stable and,
at high energy, will be the dominant stable state produced.
(If $M_6 \sim m_t~$ then, we expect, $~m_{N_6} \approx  500~ GeV$.)
$N_6\bar{N}_6~$ production will dominate the 
formation of matter in the early universe and form cold 
dark matter, as (sextet) nuclei, clumps, etc. 

Evidence for the energy scale at which 
the dominance of sextet states appears, 
may be provided by cosmic ray data.
The ``knee'' in the cosmic ray spectrum is an 
extraordinary phenomenon and suggests a 
major strong interaction change between Tevatron and LHC energies. 
A production threshold for
particles not observed at ground level would lead to an underestimation of 
energies above the threshold, and produce a knee via the pile-up of events 
below the threshold energy and a depletion of the 
spectrum above the threshold. However, a
major part of the x-section must be involved, as would be the case for the sextet
threshold. Multiple $W^{\pm}$ and $Z^0$ production will give
a huge increase of the large
$E_T$ jet cross-section implying that an unexpectedly
large fraction of shower particles will be undetected. 
There will also be
a much larger fraction of (undetected) neutrinos and, at high enough energy,
dark matter ($N_6$) production will take away a major part of the energy.
In detail, the knee suggests that, at the LHC, the new physics 
should contribute $\sim 10-20\%$ of the hadronic cross-section.
Many other effects seen in cosmic ray showers, with energies 
above the knee, also suggest new physics appears. In particular, the
production of dijets (core pairs) 
is orders of magnitude above the QCD prediction.
Also ultra high-energy events with 
$E_0 > 10^{20}~eV$ (exceeding the GZK cut-off) are not understood.
Since $N_6$'s avoid the GZK cut-off (because they are both 
neutral and massive), and have large high-energy hadronic cross-sections,
they could be reponsible. If this is the case, the mysteries of dark matter, 
the knee, and the ultra high-energy events would all have a common origin.

\section{LHC Physics} 

Via anomaly pole amplitudes, the 
hard double pomeron production of electroweak vector bosons gives jet cross-sections
comparable with normal QCD jet (non-diffractive) cross-sections, with
the boson pair cross-section estimated to be, roughly, twelve 
orders of magnitude larger than in the Standard model. 
Combining this estimate with pomeron regge theory,
gives a small transverse momentum cross-section that is 
correspondingly large. During the initial ``soft physics'' 
running period of the LHC, it should be straightforward to look for vector 
boson pairs in the CMS central detector, produced
in combination with scattered protons in the TOTEM Roman Pots.

As discussed above, there will be very large 
inclusive cross-sections for sextet states, across most of the rapidity axis. 
Multiple vector boson production will give 
jet cross-sections, at very large transverse 
momentum, that will be orders of magnitude larger than expected. 
The production cross-section for sextet nucleon pairs should also be 
hadronic in size, although stable sextet neutrons (dark matter!) may be 
difficult to detect. 
If the sextet nucleon double pomeron cross-section is 
extraordinarily large, it might be detectable in the low 
luminosity run. If not, it might be seen by the high luminosity 
detectors that will look for double pomeron production of the 
Standard Model Higgs particle. 

\section{GUT$_S$ }

The  $QCD_S$ fixed point implies that, well above the electroweak 
scale, $\alpha_s \sim \alpha_{ew} $ and so supersymmetry is not required for unification
! A priori, unification could also determine how the 
(short-distance) $SU(2)\otimes U(1)$ sextet sector 
anomaly is canceled, as well as providing an origin for masses.
Many years ago (with Kyungsik Kang) we found a remarkable, but puzzling, result. 
We looked at  asymptotically free, anomaly-free. left-handed 
unified theories that contain the sextet sector,
We discovered that a unique theory is selected, i.e.
$SU(5)$ gauge theory with the fermion representation $5+15+40+45^*$ 
($\equiv$ GUT$_S$). Amazingly,
the triplet quark and lepton sectors, which were not asked for, 
are remarkably close
to the Standard Model. There are three ``generations'' of 
quarks/anti-quarks, with quark charges $\frac{2}{3}$ and $-\frac{1}{3}$, 
and  three ``generations'' of $SU(2)$ doublet 
($SU(3)$ singlet) leptons. The puzzle is  
that the $SU(2)\otimes U(1)$ quantum 
numbers are almost, but not quite, right and there are also 
(apparently unwanted) color octet quarks with lepton-like
electroweak quantum numbers. At the time, we considered various ``anomalous
fermion phenomena'', but found no convincing dynamical route
to the Standard Model.
 
\section{ A Massless Theory of Matter ? }

In fact, GUT$_S$ has, essentially, the same infra-red and ultra-violet
properties as massless $QCD_S$. As a result the high-energy S-Matrix 
can also be constructed via reggeon diagram anomaly interactions. Although
an infra-red fixed-point keeps the $SU(5)$ coupling very small, 
reggeon infra-red divergences will
confine $SU(5)$ color in the S-Matrix. Hence, {\it all elementary
fermions will be massless and confined and the  
Dirac sea will control the dynamics,} but with 
a crucial difference from QCD$_S$. In GUT$_S$, left-handed fermion
interactions will exponentiate the initial anomalous color parity divergences 
that lead to the states and amplitudes of QCD$_S$. 
Therefore, these divergences {\it will only be produced by  
the  $SU(3)\otimes U(1)$ vector part of the theory} and will lead to
the dominance of this sector in the S-Matrix. The left-handed vector bosons,
with no SU(3) color, will aquire a mass, as described in in Section 3
and there will be a related bound-state mass spectrum in which, because of
the fermion representation structure, there will be no (unwanted) symmetries.
As yet, very little is certain, but an initial study suggests that the states and 
interactions of the Standard Model could be generated within the GUT$_S$ S-Matrix
(with the octet quarks essential for generating bound-state leptons).
If the Standard Model does, indeed, emerge in this manner it will be crucial
that (as we have seen in QCD$_S$) 
the infra-red chiral anomaly effects of a massless 
Dirac sea can produce a bound-state S-Matrix with
dramatically different properties  
from those implied, at first sight, by the underlying field theory.

\end{document}